\documentclass[11pt, letterpaper]{amsart}
\usepackage{graphicx, amssymb}

\newcommand{\myfigure}[2]{ \includegraphics*[#1]{#2} }

\newtheorem{theorem}{Theorem}

\newtheorem{prop}{Proposition}

\newtheorem{corollary}{Corollary}
\newcommand{\G}{\mathcal{G}}
\newcommand{\F}{\mathcal{F}}
\newcommand{\e}{\epsilon}

\theoremstyle{definition}
\newtheorem{example}{Example}

\newtheorem{definition}{Definition}

{\bf}{\it}
\newtheorem{remark}{Remark}
{\bf}{\it}

\title[]{On The Existence Of Consistent Price Systems}

\author[]{Erhan Bayraktar} \address{Department of Mathematics, University of Michigan} \email{erhan@umich.edu}

\author[]{Mikko S.\ Pakkanen} \address{CREATES and Department of Economics and Business, Aarhus University}
\email{msp@iki.fi}

\author[]{Hasanjan Sayit}\address{Department of Mathematics, Worcester Polytechnic Institute}
\email{hs7@WPI.EDU}

\date{\today}

\begin{document}
\maketitle

\begin{abstract} 
We formulate a sufficient condition for the existence of a consistent price system (CPS), which is weaker than the conditional full support condition (CFS). 
We use the new condition to show the existence of CPSs  for certain processes that fail to have the CFS property. In particular this condition gives sufficient conditions, under which a continuous function of a process with CFS admits a CPS, while the CFS property might be lost.

\end{abstract}
\section{Introduction}

In markets with proportional transaction costs, a \emph{consistent price system} (CPS) plays the role of  a martingale measure in both hedging and absence of arbitrage problems, as highlighted by the recent results of Guasoni, R\'asonyi, and Schachermayer (see \cite[Theorem 1.3]{Gua1} and \cite[Theorem 1.11]{Gua2}). Therefore it is crucial to study the existence of CPSs. Recall that a strictly positive adapted stochastic process $(Y_t)_{t\in [0, T]}$ defined on a filtered probability space $(\Omega, \mathcal{F}, \mathbb{F}=(\mathcal{F}_t)_{t\in [0, T]}, P)$ that satisfies the usual conditions (i.e., the filtration $\mathbb{F}$ is right continuous, and $\mathcal{F}_0$  contains all of the $P$ null sets of $\mathcal{F}$) admits an $\e $-CPS for $\e>0$ if there exists an equivalent probability measure $\tilde{P}\sim P$ and a $(\mathbb{F}, \tilde{P})$-martingale $(\tilde{Y}_t)_{t \in [0,T]}$ such that $$(1+\e)^{-1}Y_t\le \tilde{Y}_t\le (1+\e)Y_t \quad \text{ a.s. for all $t\in [0, T]$}.$$ Originally, the concept of CPS is due to Jouini and Kallal \cite{MR1338025}. See \cite{MR2030834} for further details. 
 
In \cite{Gua1}, Guasoni, R\'asonyi, and Schachermayer introduced an important condition, \emph{conditional full support} (CFS), for continuous stochastic processes and showed that CFS implies the existence of CPSs. (See equation \eqref{eq:cfs}, below, for the definition of CFS.) They proved that fractional Brownian motion (fBm) and certain continuous Markov processes possess the CFS property. Motivated by this result, in the subsequent papers  \cite{MR2432181,zanten,FEH,mikko} several other processes were shown to possess the CFS property.

In Section~\ref{sec:mainresult} of this note, we give weaker sufficient conditions for the existence of CPSs. As an application of these results, in Section~\ref{sec:applications}, we study the existence of the CPSs for transformed processes of the form $e^{f(X)}$, where $f : \mathbb{R}\rightarrow \mathbb{R}$ is a continuous function and $X$ is a continuous process with CFS. Moreover, based on these results, we construct examples of processes $f(X)$ that do not have CFS, and yet admit a CPSs.

\section{Criteria for the existence of consistent price systems}\label{sec:mainresult} 
Let us first recall the definition of random walk with retirement, introduced in \cite{Gua1}. To this end, let $(\Omega, \mathcal{G}, \mathbb{G}=(\mathcal{G}_n)_{n\geq 0}, P )$ be a discrete-time filtered probability space such that $\mathcal{G}_0 = \{\varnothing,\Omega\}$ and $\vee _n\mathcal{G}_n=\G$. 

\begin{definition}\label{rw} A \emph{random walk with retirement} is a $\mathbb{G}$-adapted process $(Z_n)_{n\geq 0}$ such that $Z_0 > 0$ and
\[
Z_n=Z_0(1+\epsilon)^{\sum_{i=1}^nR_n}, \quad n\geq 1,
\]
where $\e>0$ and $(R_n)_{n \geq 1}$ is a $\mathbb{G}$-adapted process in $\{-1, 0, 1\}$ with the following properties:
\begin{enumerate}
\item[(R1)] $R_m=0$ for all $m\geq n$ on $\{R_{n-1} =0\}$ for all $n\geq 2$;
\item[(R2)] $P(R_n=j \,| \,\G_{n-1})>0$ on $\{ R_{n-1}\neq 0\}$ for all $j\in \{-1, 0, 1\}$ and $n\geq 1$, with the convention that $\{ R_0 \neq 0\} = \Omega$;
\item[(R3)] $P(R_n\neq 0 \textrm{ for all } n\geq 1)=0$.
\end{enumerate}
\end{definition}
Any random walk with retirement $(Z_n)_{n\geq 0}$ admits an equivalent probability measure $Q\sim P$, under which it is a uniformly integrable martingale \cite[Lemma 2.6]{Gua1}. This fact will be used in our argument, below.

\begin{figure}
  \myfigure{scale=0.95}{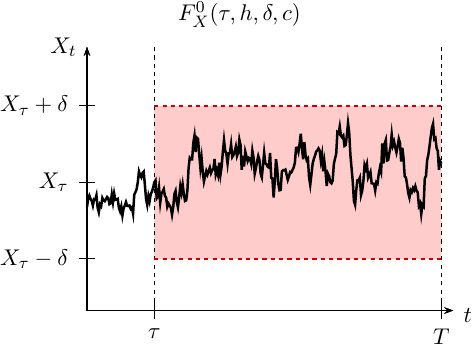}
  
  \vspace*{0.3cm}
  
  \myfigure{scale=0.95}{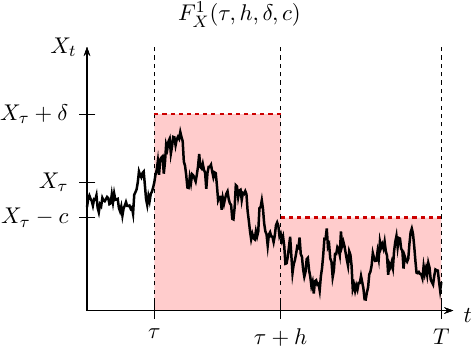}
  \myfigure{scale=0.95}{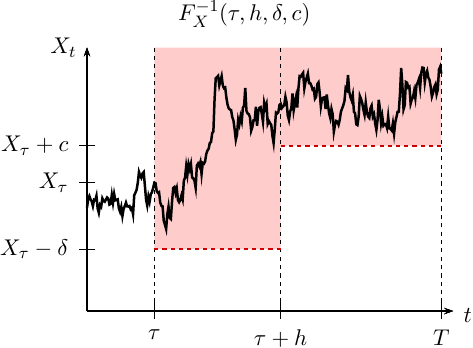}
\caption{\label{events}The events $F_X^{0}(\tau, h, \delta, c)$, $F_X^{1}(\tau, h, \delta, c)$, and $F_X^{-1}(\tau, h, \delta, c)$ in \eqref{F}.}
\end{figure}

To state our main results, let $(X_t)_{t \in [0,T]}$ be a continuous process adapted to the filtration $\mathbb{F}$. Moreover, for any $h\in (0, T)$, $\delta>0$, $c>0$, and any stopping time $\tau$ with values in $[0, T-h)$, let 
\begin{equation}\label{F}
\begin{aligned}
F_X^{0}(\tau, h, \delta, c)&:=\Big\{\sup_{t\in [0, T-\tau)}|X_{\tau+t}-X_{\tau}|<\delta \Big\}, \\
F_X^{1}(\tau, h, \delta, c)&:=\Big\{\sup_{t\in [0, h]}X_{\tau+t}<X_{\tau}+\delta \Big\}\cap \Big\{\sup_{t\in [h, T-\tau)}X_{\tau+t}<X_{\tau}-c \Big\},  \\
F_X^{-1}(\tau, h, \delta, c)&:=\Big\{\inf_{t\in [0, h]}X_{\tau+t}>X_{\tau}-\delta\Big\}\cap \Big\{\inf_{t\in [h, T-\tau)}X_{\tau+t}>X_{\tau}+c \Big\}.
\end{aligned}
\end{equation}
The event $F^0(\tau,h,\delta,c)$ is indeed independent of $h$ and $c$, but we add these arguments for consistency with $F^{-1}(\tau,h,\delta,c)$ and $F^1(\tau,h,\delta,c)$. Roughly speaking, these three events correspond to $X$ staying in a tube, moving down, and moving up, respectively, after the stopping time $\tau$---see Figure \ref{events} for an illustration. 

\begin{theorem} \label{lem1} Let $(X_t)_{t \in [0,T]}$ be a continuous process adapted to filtration $\mathbb{F}$. If there exists $\e_0>0$ such that for any $h\in (0, T)$ and stopping time $\tau$ with values in $[0, T-h)$, and $j \in \{-1, 0, 1\}$,
\begin{equation} \label{thecondition}
P\big(F_X^j(\tau, h, \log(1+\e_0), \log(1+\e_0))\big|\F_{\tau}\big)>0 \quad \textrm{a.s.,}
\end{equation}
then $(Y_t)_{t \in [0,T]}:=(e^{X_t})_{t \in [0,T]}$ admits an $\e$-CPS with $\e=(1+\e_0)^3-1$.
\end{theorem}

\begin{proof} As in proof of Theorem 1.2 of \cite{Gua1}, we set up a CPS for $Y$ using a random walk with retirement associated with $Y$. We divide the proof into three steps.

\textbf{Step 1.} Define 
\begin{eqnarray}\label{xTau}
\tau_0:=0, \quad \tau_{n}:=\inf\{t\geq \tau_{n-1}: (X_t-X_{\tau_{n-1}})\notin (-\log(1+\e_0), \log(1+\e_0)) \}\wedge T,
\end{eqnarray}
and
\begin{eqnarray}\label{xRn}
R_n:=\left\{
\begin{array}{ll}
\text{sign}(X_{\tau_n}-X_{\tau_{n-1}}),& \textrm{on $\{\tau_n<T\}$}, \\
0,&\mbox{on $\{\tau_n=T\}$}
\end{array} \right.
\end{eqnarray}
for all $n \geq 1$. Moreover, set
\begin{eqnarray}\label{rwalk}
Z_0:=Y_0, \quad Z_n:=Z_0(1+\e_0)^{\sum_{i=1}^nR_i}\quad  \textrm{for all $n\geq 1$}.
\end{eqnarray}
By construction, $\frac{1}{1+\e_0}\le \frac{Y_{\tau_n}}{Z_n}\le 1+\e_0$ for all $n\geq 0$ and $(Z_n)_{n\geq 0}$ is adapted to the filtration $(\mathcal{G}_n)_{n\geq 0}$, given by $\mathcal{G}_n=\mathcal{F}_{\tau_n}$.\\

\textbf{Step 2.} We will check that $Z$ satisfies the conditions of a random walk with retirement on the filtered probability space $(\Omega, \mathcal{G}, (\mathcal{G}_n)_{n\geq 0}, P)$, with $\mathcal{G}=\vee_{n\geq 0}\mathcal{G}_n $.
To show this, we need to check (R1)--(R3) in Definition \ref{rw}. Clearly, condition (R1) is satisfied, and (R3) is a consequence of the continuity of $X$. Therefore, we only need to check that 
\begin{eqnarray}\label{hey}
P(R_n=j|\mathcal{F}_{\tau_{n-1}})>0 \quad \mbox{on} \quad \{R_{n-1}\neq 0\}, \quad \mbox{for} \quad j \in \{-1, 0, 1\},
\end{eqnarray}
for all $n\geq 1$. This is equivalent to showing that for any $A\in \F_{\tau_{n-1}}$ with $$A\subset \{R_{n-1}\neq 0\}=\{\tau_{n-1}<T\},$$ and $P(A)>0$, $$P(A\cap \{R_n=j\})>0 \quad \text{for all}\; j \in \{-1, 0, 1\}.$$
Let $s<T$ be such that $P(A\cap \{\tau_{n-1}<s\})>0$. Let $B=A\cap \{\tau_{n-1}<s\}$ and $h=\frac{T-s}{4}$. Denote $$\tau_{n-1}^B=\tau_{n-1}1_{B}+\frac{T+s}{2} 1_{\Omega\backslash B}.$$ Note that $\tau_{n-1}^B$ is a stopping time and its values are in $[0, T-h)=[0, \frac{T+s}{2}+\frac{T-s}{4})$. By the assumption of the theorem, we have $$P\left(F_X^j(\tau_{n-1}^B, h, \log(1+\e_0), \log(1+\e_0)))\big|\F_{\tau_{n-1}^B}\right)>0  \quad \text{a.s.},$$ for any $j \in \{-1, 0, 1\}$. Note that $B\in \mathcal{F}_{\tau_{n-1}^B}$  with $P(B)>0$, and therefore, the events $$B\cap F_X^j(\tau_{n-1}^B, h, \log(1+\e_0), \log(1+\e_0)), \quad z\in \{-1,0,1\}$$ have positive probability, which, in turn, implies  $P(\{R_n=j\}\cap B)>0$ for any $j \in \{-1, 0, 1\}$. Since $B\subset A$, the result follows. \\

\textbf{Step 3.} Since $(Z_n)$ is a random walk with retirement, thanks to Lemma 2.6 of \cite{Gua1}, there exists an equivalent probability measure $Q\sim P$ such that $(Z_n, \mathcal{G}_n)_{n\geq 0}$ is a uniformly integrable martingale. Let $Z_{\infty}=\lim_{n \to \infty} Z_n$. For each $t\in [0, T]$, set $\tilde{Z}_t=E_{Q}[Z_{\infty}|\mathcal{F}_t]$. Observe that $\tilde{Z}_{\tau_n}=E_Q[Z_{\infty}|\mathcal{F}_{\tau_n}]=Z_n$, and that 
$\tilde{Z}_t=E_Q[\tilde{Z}_{\tau_n}|\mathcal{F}_t]$ on the set $\{\tau_{n-1}\le t \le \tau_n\}$ for all $n\geq 0$. Thus the following holds
\begin{eqnarray}\label{nine}
\frac{\tilde{Z}_t}{Y_t}1_{\{\tau_{n-1}\le t \le \tau_n\}}=E_Q\left[\frac{Z_n}{Y_t}1_{\{\tau_{n-1}\le t \le \tau_n\}}\bigg|\mathcal{F}_t\right], \; \;  n\geq 1.
\end{eqnarray}
We write $\frac{Z_n}{Y_t}=\frac{Z_n}{Y_{\tau_n}}\frac{Y_{\tau_{n-1}}}{Y_t}\frac{Y_{\tau_{n}}}{Y_{\tau_{n-1}}}$. Note that each of $\frac{Z_n}{Y_{\tau_n}}$, $\frac{Y_{\tau_{n-1}}}{Y_t}$, and $\frac{Y_{\tau_{n}}}{Y_{\tau_{n-1}}}$ takes values in 
$((1+\e_0)^{-1}, 1+\e_0)$ on the set $\{\tau_{n-1}\le t\le \tau_n\}$. Therefore, from (\ref{nine}), we have $$(1+\e_0)^{-3} \le \frac{\tilde{Z}_t}{Y_t}\le (1+\e_0)^3 \quad \text{on}  \; \{\tau_{n-1}\le t\le \tau_n\}.$$ Since $\cup_{n=1}^{\infty}\{\tau_{n-1}\le t\le \tau_n\}=\Omega$, we conclude that
\begin{eqnarray}\label{three}
(1+\e_0)^{-3}\le \frac{\tilde{Z}_t}{Y_t} \le (1+\e_0)^3. 
\end{eqnarray}
Therefore $\tilde{Z}_t$ is an $\e$-CPS for $Y_t$, with $\e=(1+\e_0)^3-1$. 
\end{proof}
\begin{remark}\label{rem1.5} If $X$ is adapted to a sub-filtration  $\mathbb{F}'$ of $\mathbb{F}$ and (\ref{thecondition}) holds with respect to $\mathbb{F}$ for $\e_0>0$, then it also holds with respect to the smaller filtration $\mathbb{F}'$ for $\e_0$. 
\end{remark}

The condition \eqref{thecondition} in Theorem~\ref{lem1} needs, of course,  to be checked for a very wide class of stopping times. Depending on the process $X$, direct verification of \eqref{thecondition} might be a difficult task. To overcome this difficulty, we establish the following variant of Theorem~\ref{lem1} with a sufficient condition that involves only \emph{deterministic} times. 

\begin{theorem} \label{thm:nostoppingtimes} Let $(X_t)_{t \in [0,T]}$ be a continuous process adapted to filtration $\mathbb{F}$. If there exists $\gamma>0$  such that for any $h\in (0, T)$, $t \in [0, T-h)$, $\delta \in (0,\gamma)$, $c\in (0,\gamma)$, and $j \in \{-1, 0, 1\}$,
\begin{equation} \label{nostopcondition}
P\big(F_X^j(t, h, \delta, c)\big|\F_t\big)>0 \quad \textrm{a.s.,}
\end{equation}
then $(Y_t)_{t \in [0,T]}:=(e^{X_t})_{t \in [0,T]}$ admits an $\e$-CPS for any $\e>0$.
\end{theorem}

\begin{proof}
By Theorem~\ref{lem1}, it suffices to show that \eqref{nostopcondition} holds whenever $t$ is replaced with any stopping time $\tau$ that assumes values in $[0, T-h)$. We use a strategy that is similar to the proof of Lemma 2.9 of \cite{Gua1} and assume, contrapositively, that there is $h \in (0,T)$, stopping time $\tau$ with values in $[0,T-h)$, $\delta\in(0,\gamma)$, and $c\in(0,\gamma)$ such that
\begin{equation}\label{eq:contraposition}
P(A) >0, \quad \textrm{where} \quad A := \Big\{ P \big(F^j_X(\tau,h,\delta,c)\big|\mathcal{F}_\tau\big) =0 \Big\}, 
\end{equation}
for some $j \in \{-1,0,1 \}$. We will consider here only the case $j=-1$.  When $j=0$, it suffices to invoke Lemma 2.2 of \cite{FEH}, whereas the case $j=1$ is completely analogous to $j=-1$.

For brevity, let us write $B := \Omega \setminus F^{-1}_X(\tau,h,\delta,c)$. By \eqref{eq:contraposition} and the definition of conditional expectation, we have $1_A 1_B = 1_A$. The continuity of the paths of $X$ implies that $A = \cup_{q \in \mathbb{Q}} A_q$, where 
\begin{equation*}
A_q := A \cap \bigg\{ q-\frac{h}{2}\leq \tau \leq q \bigg\} \cap \bigg\{ \sup_{t \in [\tau,q]} (X_\tau- X_t) \leq \frac{\min(\delta, \gamma-c)}{2} \bigg\} \in \mathcal{F}_q.
\end{equation*}
Since $P(A)>0$, there is $q \in \mathbb{Q}$ such that $P(A_q)>0$. Let us consider the stopping time
\begin{equation*}
\rho := \inf \big\{ t > \tau : X_\rho - X_\tau \leq -\delta 1_{\{L_\rho < h \}} + c 1_{\{L_\rho \geq h \}}\big\},
\end{equation*}
where $L_t := \int_0^t 1_{[\tau,T]}(s) ds$,~$t \in [0,T]$, that clearly satisfies $X_\rho - X_\tau \leq -\delta$ on $A \cap B\cap \{\rho < \tau +h \}$ and $X_\rho - X_\tau \leq c$ on $A \cap B \cap \{\rho \geq \tau +h \}$.
Note that $q+h/2 \leq \tau + h < T$ and $\rho > q$ on $A_q$, and that $A_q \cap B \cap \{  \rho < q+h/2\} \subset A \cap B \cap \{ \rho < \tau + h\}$. Hence, on $A_q \cap B \cap \{  \rho < q+h/2\}$,
\begin{equation*}
X_\rho - X_q = X_\rho - X_\tau + X_\tau - X_q \leq X_\rho - X_\tau + \frac{\delta}{2} \leq -\frac{\delta}{2}.
\end{equation*}
Moreover, on $A_q \cap B \cap \{  \rho \geq q+h/2\}\subset A$,
\begin{equation*}
X_\rho - X_q \leq X_\rho - X_\tau + \frac{\gamma-c}{2} \leq \frac{\gamma+c}{2}.
\end{equation*}
We have thus shown that $A_q \cap B \subset A_q \cap C$, where
\begin{equation*}
C := \Omega \setminus F^{-1}_X\bigg(q,\frac{h}{2},\frac{\delta}{2},\frac{\gamma+c}{2}\bigg).
\end{equation*}
Furthermore, $\delta/2 \in (0,\gamma)$ and $(\gamma+c)/2 \in (0,\gamma)$.
Finally, we have
\begin{equation*}
P(A_q) \geq E[ 1_{A_q}P(C | \mathcal{F}_q) ] = E[1_{A_q}1_C] \geq E[1_{A_q}1_B]   = E[1_{A_q}1_A1_B] = P(A_q),
\end{equation*}
whence $P(C | \mathcal{F}_q) = 1$ on $A_q$, and the assertion follows.
\end{proof}

Let us briefly compare the criteria above to the conditional full support property, mentioned in the introduction. Recall that a 
 continuous process $(X_t)_{t\in [0, T]}$ has \emph{conditional full support} (CFS) if 
\begin{equation}\label{eq:cfs}
\mbox{supp\,Law}(X_{\theta}; t\le \theta \le T|\mathcal{F}_{t})=C_{X_t}[t, T] \quad \textrm{a.s.,}
\end{equation}
where $C_x[t, T]$ denotes the space of continuous functions $[t, T]\rightarrow \mathbb{R}$ with $f(t)=x$ and ``supp" denotes the support (the smallest closed set of probability one). Actually, the CFS property holds if and only if \eqref{eq:cfs} is satisfied also when $t$ is replaced with an arbitrary stopping time \cite[Lemma 2.9]{Gua1}.

The sufficient conditions for the existence of an $\epsilon$-CPS for arbitrarily small $\epsilon>0$ established in Theorems \ref{lem1} and \ref{thm:nostoppingtimes} are weaker than CFS. In particular, they are \emph{local} in the sense that they do not require that $X$ remains $\e$-close to, e.g., a continuous function with arbitrarily large maximum with positive conditional probability, like CFS does. This is illustrated by the following consequence of Theorem \ref{thm:nostoppingtimes}. 

\begin{corollary}\label{cor:support}
Let $(X_t)_{t \in [0,T]}$ be a continuous process adapted to filtration $\mathbb{F}$. If there exists $\delta>0$ such that for any $t \in [0,T)$ and continuous, monotone function $g : [t,T] \rightarrow [-\delta,\delta]$ with $g(t)=0$,
\begin{equation}\label{eq:monsupport}
g + X_t \in \mathrm{supp}\,\mathrm{Law}(X_{\theta}; t\le \theta \le T|\mathcal{F}_{t}) \quad \textrm{a.s.,}
\end{equation}
then $(Y_t)_{t \in [0,T]}:=(e^{X_t})_{t \in [0,T]}$ admits an $\e$-CPS for any $\e>0$.
\end{corollary}

\begin{proof}
It suffices to note that \eqref{eq:monsupport} is equivalent to the requirement that for any $\delta>0$,
\begin{equation*}
P\bigg( \sup_{\theta \in [t,T]} |X_\theta-X_t - g(\theta)|<\delta \,\bigg|\, \mathcal{F}_t\bigg) >0 \quad \textrm{a.s.,} 
\end{equation*}
and then apply Theorem \ref{thm:nostoppingtimes}.
\end{proof}

\section{Application to transformed processes}\label{sec:applications}

As an application of the results above, we study the existence of CPSs for processes of the form $e^{f(X)}$, where $f : \mathbb{R} \rightarrow \mathbb{R}$ is a continuous surjection and $X$ is a process with CFS.

\begin{prop}\label{lm2} Assume that $X$ is continuous process with CFS. Let $\delta_0>0$ and $f : \mathbb{R} \rightarrow \mathbb{R}$ be a continuous function that satisfies either of the following:
\begin{eqnarray*}
\begin{array}{ll}
(a)   \lim_{x\rightarrow -\infty}f(x)=-\infty, \quad \lim_{x\rightarrow +\infty}f(x)=+\infty,  \quad \text{and} \quad  \min_{y\geq x}(f(y)-f(x))>-\delta_0; \\
(b) \lim_{x\rightarrow -\infty}f(x)=+\infty, \quad \lim_{x\rightarrow +\infty}f(x)=-\infty, \quad \text{and} \quad \max_{y\geq x}(f(y)-f(x))<\delta_0.
\end{array}
\end{eqnarray*}
Then 
\begin{eqnarray}\label{fff}
P\left(F_{f(X)}^j(\tau, h, \delta_0, c_0)\big|\F_{\tau}\right)>0, \; \; j \in \{-1, 0, 1\}, 
\end{eqnarray}
for any $h\in (0, T)$, any $\mathbb{F}$ stopping time $\tau$ with values in $[0, T-h)$, and any $c_0>0$. 
\end{prop}

\begin{proof} We will show the result for continuous functions $f$ that satisfy condition (a). The proof for any $f$ that satisfies condition (b) is similar and will be omitted. 

 Let $h\in (0, T)$ and $\tau$ be an $\mathbb{F}$-stopping time with values in $[0, T-h)$. In order to prove \eqref{fff}, we need to show that $P(A\cap F_{f(X)}^j(\tau, h, \delta_0, c_0))>0$ for any $A\in \F_{\tau}$ with $P(A)>0$. Fix any $A\in \F_{\tau}$ with $P(A)>0$. Let $k>0$ be such that the event $$B=A\cap \{-k<X_{\tau}<k\}\cap \{-k<f(X_{\tau})<k\}$$ has positive probability. Note that $B\in \F_{\tau}$. Since $f$ is uniformly continuous on $[-k-1, k+1]$, there exists $\delta \in [0,1]$ such that $|f(y)-f(x)|< \delta_0$, whenever $x, y\in [-k-1, k+1]$ and $|x-y|<\delta$.

\textbf{(i)} Proof that $P(A\cap F_{f(X)}^0(\tau, h, \delta_0, c_0))>0$ :  Note that $$\sup_{t\in [0, T-\tau)}|f(X_{\tau+t})-f(X_{\tau})|<\delta_0 \quad \text{on}\;B\cap F_X^0(\tau, h, \delta, c),$$ for any  $c>0$, and by our assumption, we have that
$P(B\cap F_X^0(\tau, h, \delta, c))>0$ . Therefore, $P(B\cap F_{f(X)}^0(\tau, h, \delta_0, c_0))>0$, which implies $P(A\cap F_{f(X)}^0(\tau, h, \delta_0, c_0))>0$. 

\textbf{(ii)} Proof that $P(A\cap F_{f(X)}^{1}(\tau, h, \delta_0, c_0))>0$: Let $\tilde{c}>0$ be such that 
$f(x)<-c_0-k$ for all $x<-\tilde{c}$. By our assumption on $X$, we have that $P(F_X^{1}(\tau, h, \delta, \tilde{c}+k)|\F_{\tau})>0$ a.s. Therefore, $P(B\cap F_X^{1}(\tau, h, \delta, \tilde{c}+k))>0$. Observe that on $B\cap F_X^{1}(\tau, h, \delta, \tilde{c}+k)$, $$\sup_{t\in [0, h]}(X_{\tau+t}-X_{\tau})<\delta \quad \text{and}\quad X_{\tau}\in (-k, k).$$ Therefore, if $X_{\tau+t}\geq X_{\tau}$, then $0\le X_{\tau+t}-X_{\tau}\le \delta \in [0,1]$, which implies that $$X_{\tau},  X_{\tau+t}\in [-k-1, k+1].$$ As a result, $f(X_{\tau+t})-f(X_{\tau})<\delta_0$. If, on the other hand, $X_{\tau+t}\le X_{\tau}$, then since $\sup_{y\geq x}(f(x)-f(y))<\delta_0$, we have $f(X_{\tau+t})-f(X_{\tau})<\delta_0$. Therefore, on $B\cap F_X^{1}(\tau, h, \delta, \tilde{c}+k)$, $$\sup_{t\in [0, h]}(f(X_{\tau+t})-f(X_{\tau}))<\delta_0.$$ 

Moreover, on $B\cap F_X^{1}(\tau, h, \delta, \tilde{c}+k)$, we have that $$\sup_{t\in [h, T-\tau)}(X_{\tau+t}-X_{\tau})<-\tilde{c}-k \quad \text{and} \quad X_{\tau}\in (-k, k).$$ This implies that $$\sup_{t\in [h, T-\tau)}X_{\tau+t}<-\tilde{c},$$ which in turn implies that  $$\sup_{t\in [h, T-\tau)}f(X_{\tau+t})<-c_0-k \quad \text{on}\;B\cap F_X^{1}(\tau, h, \delta, \tilde{c}+k).$$ Now, since $f(X_{\tau})\in (-k, k)$ on $B\cap F_X^{1}(\tau, h, \delta, \tilde{c}+k)$, it follows that $$\sup_{t\in [h, T-\tau)}(f(X_{\tau+t})-f(X_{\tau}))<-c_0 \quad \text{on}\quad B\cap F_X^{1}(\tau, h, \delta, \tilde{c}+k).$$ We conclude that $P(B\cap F_{f(X)}^{+}(\tau, h, \delta_0, c_0))>0$ from which the result follows since $B \subset A$.

\textbf{(iii)} Proof that $P(A\cap F_{f(X)}^{-1}(\tau, h, \delta_0, c_0))>0$: The proof is similar to part (ii).
\end{proof}

 The properties (a) and (b) above essentially mean ``nearly increasing'' and ``nearly decreasing'' respectively, and they would reduce to monotonicity for $\delta_0=0$. In fact, for this particular case the following holds true.

\begin{corollary}\label{cor:incso}
 If $X$ is a continuous process with CFS and $f: \mathbb{R} \rightarrow \mathbb{R}$ is a monotone, continuous surjection, then $(Y_t)_{t \in [0,T]}=(e^{f(X_t)})_{t \in [0,T]}$ admits an $\e$-CPS for any $\e>0$.
\end{corollary} 
\begin{proof} Assume $f$ is non-decreasing and satisfies the first two conditions of (a) in Proposition~\ref{lm2}. Then it also satisfies the  third condition of (a) for any $\delta_0>0$. Therefore, by Proposition~\ref{lm2},  (\ref{fff}) holds for any $\delta_0>0$ and $c_0>0$. Thus, from Theorem~\ref{lem1}, we  conclude that $Y_t$ admits $\e$-CPS for any $\e>0$. The proof for the case of non-increasing function
follows similarly. 
\end{proof}
It is worth noting that unless the continuous functions are strictly monotonous in the above corollary, $f(X)$ does not have CFS in general. The next corollary covers cases when the continuous function $f$ is not monotonous.

\begin{corollary}\label{prop2} Let $X$ be a continuous process with CFS. If $f : \mathbb{R} \rightarrow \mathbb{R}$ is a continuous function that satisfies the first two conditions in either (a) or (b) in Proposition~\ref{lm2}, then for any $\delta_0>0$ we can find a small enough $\alpha>0$ such that $g(x):=\alpha f(x)$ satisfies
\begin{eqnarray}\label{ffg}
P\left(F_{g(X)}^j(\tau, h, \delta_0, H)\big|\F_{\tau}\right)>0, \; \; j \in \{-1, 0, 1\}, 
\end{eqnarray}
for any $h\in (0, T)$, any $\mathbb{F}$ stopping time $\tau$ with values in $[0, T-h)$, and any $H>0$. 
In particular,
\begin{enumerate}
\item[(a)] If $f$ satisfies the first two conditions in (a) of Proposition~\ref{lm2} and $d:=\min_{y\geq x}(f(y)-f(x))<0$, we can let $\alpha$ to be any number in
$\left(0, \frac{\delta_0}{|d|}\right)$. 
\item[(b)] If $f$ satisfies the first two conditions in (b) of Proposition~\ref{lm2} and $d_0=\max_{y\geq x}(f(y)-f(x))>0$, we can let $\alpha$ to be any number in
$\left(0, \frac{\delta_0}{d_0}\right)$.  
\end{enumerate} 
\end{corollary}

\begin{remark}
It is clear that the surjectivity of $f$ is not a \emph{necessary} condition for the existence of CPSs for $Y= e^{f(X)}$. In particular, if $f: \mathbb{R} \rightarrow (a,b)$ is a bijection, where $-\infty < a < b < \infty$ and $X$ has CFS, then the results of \cite{FEH} can be used to construct CPSs for $Y$. However, when $f$ is not bijective and assumes the value $a$ or $b$, it appears to be an open problem whether $Y$ can have a CPS.
\end{remark}

The next example is to illustrate how Corollary \ref{prop2} can be applied.

\begin{example}\label{yt} Consider the process  $$Y^{(\alpha)}_t=\exp\left[\alpha[(B_t^H)^3+(B_t^H)^2]\right],\quad t \in [0,T],$$ where $B^H$ is a fractional Brownian motion with Hurst parameter $H \in (0,1)$. The function $f(x)=x^3+x^2$ satisfies the first two conditions in (a) of Proposition~\ref{lm2}. Also, $$d=\min_{y\geq x}(f(y)-f(x))=-\frac{12}{27}.$$ Therefore, for any $\delta_0>0$ the process $Y^{(\alpha)}$ admits an $(e^{3\delta_0}-1)-$CPS 
 whenever $\alpha\in \left[0, \frac{27}{12}\delta_0\right)$.
\end{example}

The following is an important example where $X$ has CFS and $f(X)$ does not, while $e^{f(X)}$ admits CPSs.
\begin{example} \label{ex3}  First, let us recall an implication of the CFS property:  If $X$ has a CFS, then
\begin{eqnarray}\label{scfs}
P\left(A\cap\left\{\sup_{t\in[0, T-\tau]}|X_{\tau+t}-(X_{\tau}+f(t))|<\e\right\}\right)>0,
\end{eqnarray} 
for any $[0, T]$ valued stopping time $\tau$, and any $A\in \mathcal{F}_{\tau}$ with $P(A)>0$, and any $\e>0$ and $f\in C[0, T]$. (As mentioned before, this follows from Lemma 2.9 of \cite{Gua1}.)

Now, let $(B_t)_{t \in [0,1]}$ be a standard Brownian motion. For $\alpha>0$, consider $(S^{(\alpha)}_t)_{t \in [0,1]}=(\alpha f(B_t))_{t \in [0,1]}$, where
\begin{eqnarray*}
f(x)=\left\{
\begin{array}{ll} 
|x|,& x\geq -1,\\
x+2,& x< -1.
\end{array}
\right.
\end{eqnarray*}
Let us prove that $S^{(\alpha)}$ does not have the CFS property for any $\alpha \in [0,1]$. 
Let $$\tau:=\inf\{t\geq 0: |B_t|=1\}\wedge 1.$$ On the set $\{\tau=1\}$ the paths of the process $f(B)$ are non-negative, whereas on $\{\tau<1\}$ we have that $\sup_{t\in [0, 1]}f(B_t) \geq 1$. Therefore, if we let $g(t)=-t$, then we have $$P\left(\sup_{t\in [0, 1]}|S_t^{(\alpha)}-S_0^{(\alpha)}-g(t)|\geq \alpha\right)=1.$$ Thus, $S^{(\alpha)}$ does not have the CFS property for any $\alpha \in [0,1]$.

On the other hand, $$d=\inf_{y\geq x}(f(y)-f(x))=-1.$$  For any $\delta_0>0$ the process $e^{\alpha f(B)}$ admits an $(e^{3\delta_0}-1)-$CPS, for all $\alpha\in (0, \delta_0)$, thanks to Corollary~\ref{prop2}.

Finally, it is worth stressing that without transaction costs, $S^{(\alpha)}$ does admit arbitrage opportunities. It follows from the CFS property of Brownian motion that the simple short strategy $-1_{(\tau_1,\tau_2]}$, where
\begin{equation*}
\tau_1 := \inf \Big\{t \geq 0 : S^{(\alpha)}_t = \alpha, \, \min_{s \in [0,t]} S^{(\alpha)}_t < 0 \Big\} \wedge 1, \quad \tau_2 := \inf \big\{ t \geq \tau_1 : S^{(\alpha)}_t = 0\big\} \wedge 1,
\end{equation*}
is an arbitrage.
\end{example}

\section*{Acknowledgements} 

We are grateful to the anonymous referee for his/her comments which helped us improve our paper. We would like to express our thanks to Paolo Guasoni for his helpful comments. E. Bayraktar is supported in part by the National Science Foundation under a Career grant, DMS-DMS-0955463, and in part by the Susan M. Smith Professorship. M.~S.~Pakkanen acknowledges support from CREATES, funded by the Danish National Research Foundation, and from the Aarhus University Research Foundation regarding the  project ``Stochastic and Econometric Analysis of Commodity Markets".

\bibliography{references-edit}
\bibliographystyle{plain}
\end{document}